\documentclass[preprint,12pt,3p]{elsarticle}
\usepackage{wrapfig}
\usepackage{subcaption}
\usepackage{multicol}
\usepackage{hyperref}
\begin{document}
\begin{frontmatter}
\title {Understanding of different types of rolling motion of $C_{60}$ on aluminium surface.}
\tnotetext[label0]{}
\author{Sourabh Kumar*}
\author{Aniruddha Chakraborty}
\address{School of Basic Sciences, Indian Institute of Technology Mandi, Kamand, Himachal Pradesh 175075, India.}
\begin{abstract}
Can a $C_{60}$ molecule chemisorbed on an aluminium surface, exhibit rolling motion and even move on the surface? Here we presents the results of our calculation which shows that $C_{60}$ can move on that surface utilizing its own rolling motion. This rolling motion involves breaking and making of several $Al-C$ bonds. We have analyzed three possible mechanisms, in one, only carbon atoms of hexagonal rings are involved in the process, in another case only carbon atoms of pentagonal rings are involved in the process  and in the last case carbon atoms of both the hexagonal and pentagonal rings are involved in the process. We found that the activation energy for rolling motion involving, only hexagonal rings is $1.69$ kcal mol$^{-1}$, only pentagonal rings is $5.97$ kcal mol$^{-1}$ and both hexagonal and pentagonal rings is $121.38$ kcal mol$^{-1}$. Rate constants are estimated using classical version of transition state theory for rolling motion involving, only hexagonal rings is $3.45 \times 10^{13} sec^{-1}$, only pentagonal rings is $17.01 \times 10^{-79} sec^{-1}$ and both hexagonal and pentagonal rings is $28.92 \times 10^{2} sec^{-1}$.
\end{abstract}
\end{frontmatter}

\section{Introduction}
%\label{sec1}
%{2}
\noindent In the recent past, design and synthesis of molecular machines has got considerable attention from the research community \cite{ACC}. Those molecules, which can perform any type of unusual motion, {\it e.g.}, rotation, shuttling, harpooning  are being investigated \cite{Balzani}. In this letter we report the results of one such investigation. We put one $C_{60}$ molecule on the aluminium surface and explore the possibilities of rolling motion of the $C_{60}$ molecule on that aluminium surface \cite{roll}.

\section{Computational Details}
\noindent The molecular mechanics calculations were carried out using the MM+ force field Incorporated in Hyperchem version 7.0, with Polak-Ribiere gradient algorithm and a root mean square gradient of 0.001 kcal/(\AA mol).
%
%{2}
%\includegraphics[scale=.5]{n1.png}
%\caption\small{Fig. 1. $C_{60}$}
%
{2}
\section{Absorption of Fullerene molecule on aluminium surface}
\begin{figure}[ht]
\centering
\includegraphics[scale=.23]{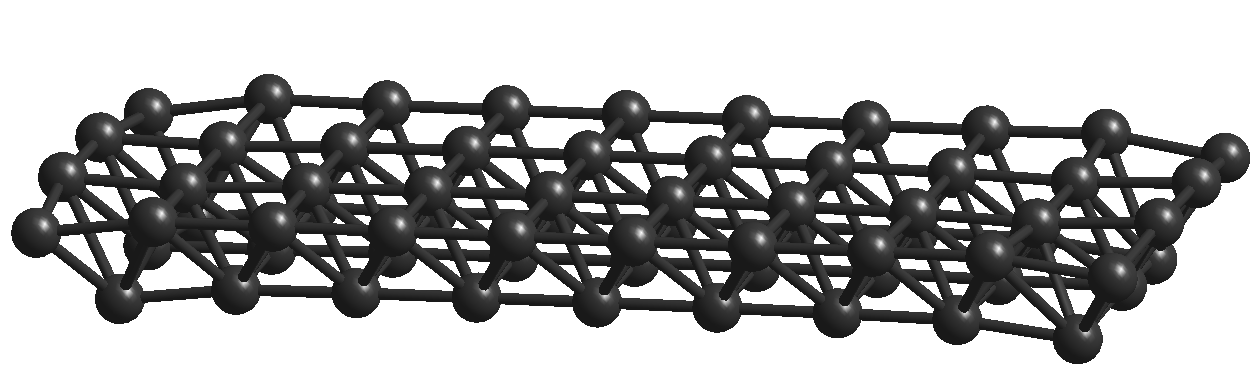}  
\caption{Al lattice made by using $67$ $Al$ atoms.}
\end{figure}
\noindent We start with a fcc (face centered cubic) lattice made by using $67$ $Al$ atoms. Surface of this lattice in general provides cubic absorption sites. Molecular mechanics method is used to optimize the structure of this Al lattice and the energy this optimized lattice is found to be $821.44$  kcal mol$^{-1}$. The average distance between two aluminium atoms in this optimized lattice is found to be $2.36$\AA. Then we place one $C_{60}$ molecule on the surface of this optimized Al lattice and we optimize the geometry of the whole system {\it i.e.}, $C_{60}$ molecule seating on the surface of the Al lattice. Due to absorption of $C_{60}$ molecule on this $Al$ surface, covalent bonds are formed between the carbon atoms of the $C_{60}$ molecule and the aluminium atoms of the Al lattice. As a result of these four new bonds, the shape of this $C_{60}$ molecule sitting on the Al surface, gets distorted as shown in Fig.2. In Fig.2 only two new C-Al bonds in the front side can be seen but there are two more C-Al bonds which are not visible.  
\begin{figure}[ht]
\centering
\includegraphics[scale=.30]{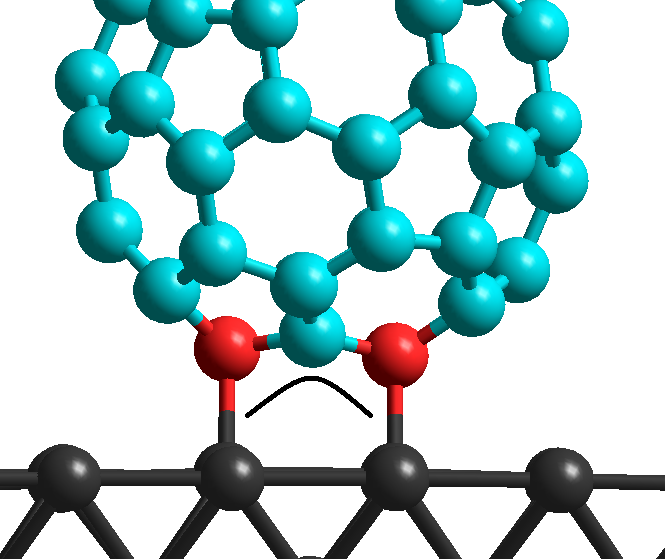}
\caption{Distorted $C_{60}$ molecule on the Aluminium surface. One curve is used to indicate those bonds which are under strain.}
\end{figure}
\noindent Therefore four carbon atoms of the $C_{60}$ molecule are participating in bond formation with the surface of the Aluminium lattice - these four carbon atoms are present either in pentagonal ring or in hexagonal ring. Some hexagonal rings (case: A) surrounded only by hexagonal rings while some other hexagonal rings (case: B) are surrounded by both pentagonal and hexagonal rings and all pentagonal rings are always surrounded by hexagonal rings only (case: C). In Table. 1, a summary of optimized energy value for all three cases are given. The average value of the Al-C bond length is found to be 1.98\AA. 

\begin{figure}[ht]
\centering
\begin{subfigure} {0.3\textwidth}
\includegraphics[scale=.35]{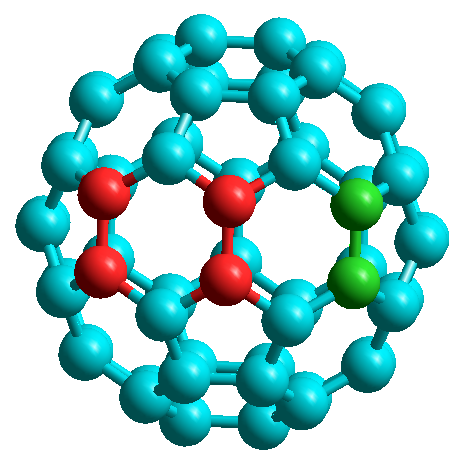}
\caption{This figure shows one hexagonal ring which is surrounded only by hexagonal rings.}
\end{subfigure}
\begin{subfigure} {0.3\textwidth}
\includegraphics[scale=.35]{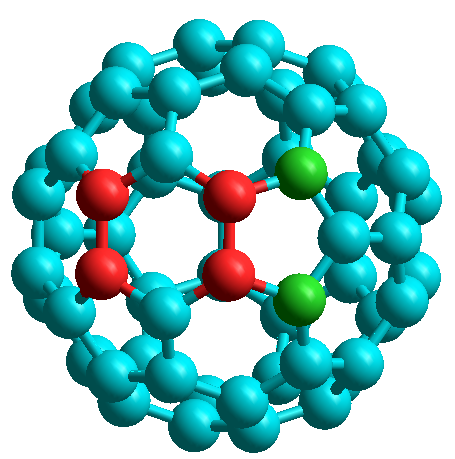}
\caption{This figure shows one hexagonal ring which is surrounded by both pentagonal and hexagonal rings.}
\end{subfigure}
\begin{subfigure} {0.3\textwidth}
\includegraphics[scale=.35]{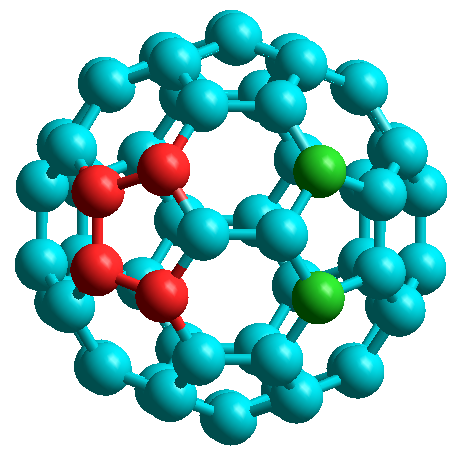}
\caption{This figure shows one pentagonal ring which is surrounded by hexagonal rings only.}
\end{subfigure}
\caption{Different types of carbon rings in $C_{60}$ molecule.}
\end{figure}

\begin{table}
\caption{Optimized energy of $C_{60}$ molecule adsorbed on the surface of aluminium lattice.}
\begin{center}
\begin{tabular}{|c|c|}
\hline
Adsorbed & Optimized Energy  \\
     $C_{60}$   &  (kcal mol$^{-1}$)  \\
\hline
Case A  & 1109.55\\
\hline
Case B & 1109.60 \\
\hline
Case C & 1090.18 \\
\hline
\end{tabular}
\end{center}
\end{table}

%\vspace{.60cm}

%\begin{tabular}{|c|c|}
%\hline
%Adsorbed & Optimized Energy  \\
%     $C_{60}$   &  (kcal mol$^{-1}$)  \\
%\hline
%on 39 Al atoms cluster  & 1063.28\\
%\hline
%on 46 Al atoms cluster  & 1077.66 \\
%\hline
%on 53 Al atoms cluster & 1087.81 \\
%\hline
%on 60 Al atoms cluster & 1098.55 \\
%\hline
%\end{tabular}

%\vspace{.50cm}

\begin{figure}[ht]
\centering
\begin{subfigure} {.3\textwidth}
\includegraphics[scale=.28]{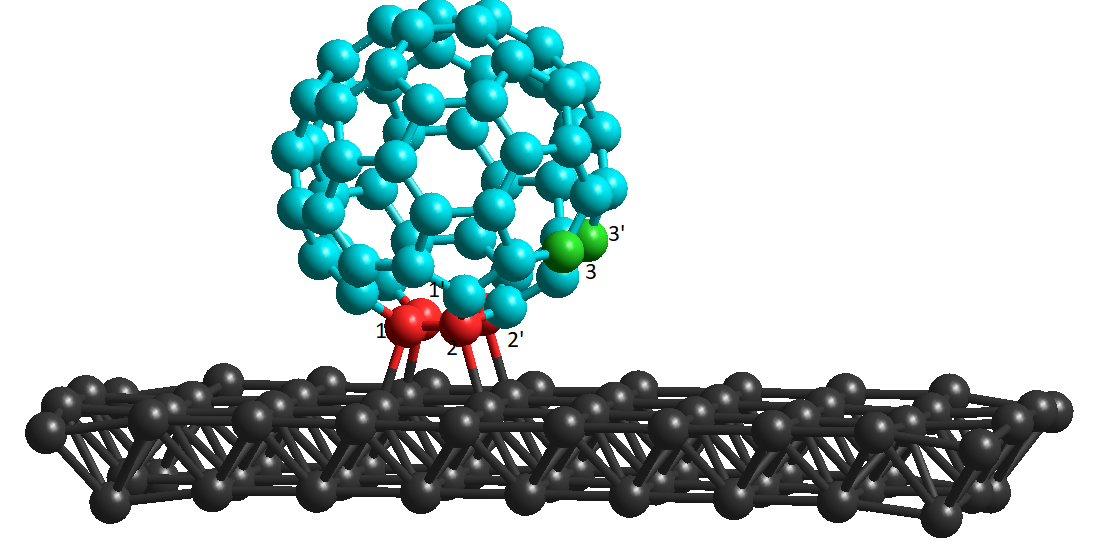}
\caption{Initial State.}
\end{subfigure}
\begin{subfigure}{.3\textwidth}
\includegraphics[scale=.28]{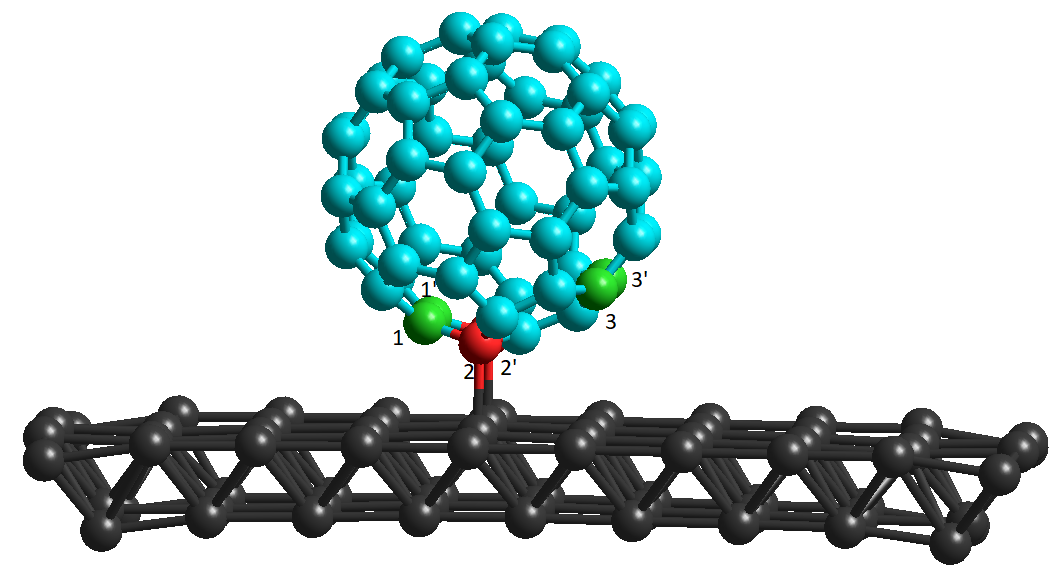} 
\caption{Intermediate State.}
\end{subfigure}
\begin{subfigure}{.3\textwidth}
\includegraphics[scale=.28]{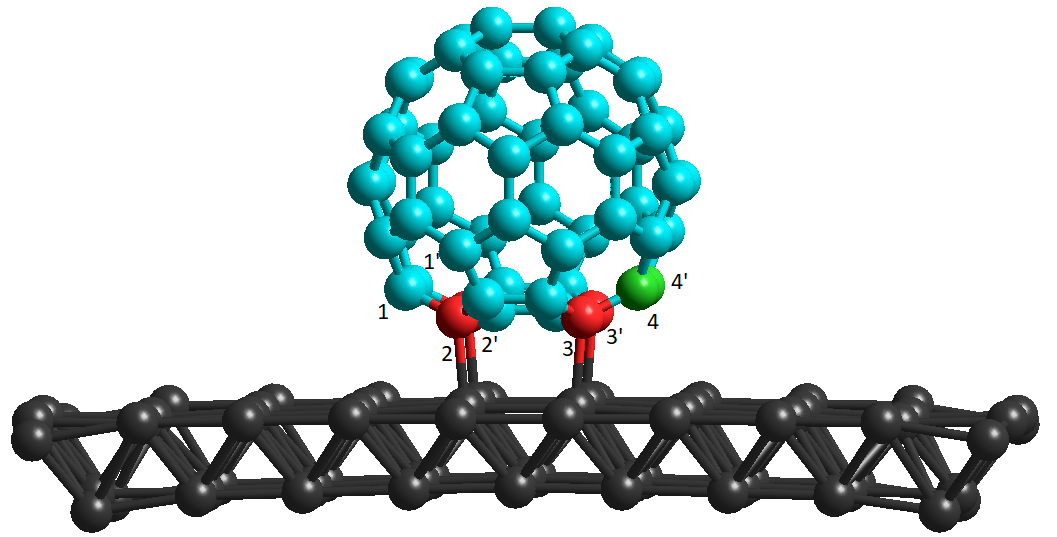} 
\caption{Final State.}
\end{subfigure}
\caption{Possibility of rolling motion of adsorbed $C_{60}$ on Al surface}
\end{figure}

\section{Rolling motion of fullerene on aluminium-surface:}

\noindent In this section we consider the possibility of rolling motion of $C_{60}$ on the surface of Aluminium lattice. We do not consider the sliding motion because, in general the activation energy for sliding motion is quite larger than the rolling motion \cite{bidisa}. We use an interesting approach to estimate the activation energy for rolling motion of $C_{60}$ molecule on the surface of Al-lattice. We first optimize the geometry of the whole system ($C_{60}$ plus Al- lattice) with $C_{60}$ molecule in the initial state and then we calculate energy by increasing the length of those two C-Al bonds which are going to be broken as a result of rolling motion of $C_{60}$. Then we repeat the same calculation  by optimizing the geometry of the whole system ($C_{60}$ plus Al- lattice) with $C_{60}$ molecule in the final state and then we calculate energy by decreasing the length of those two C-Al bonds which are going to be formed as a result of rolling motion of $C_{60}$. Then we find when (value of four C-Al bond distances which are effected by rolling motion) the energy of initial state is equal or more than the energy of the final state - that geometry corresponds to the Transition state. For checking the accuracy of our method, we use the same procedure for estimating activation energy for several systems and find that the our results are in good agreement with those available in literature \cite{bidisa}.  As we have four carbon atoms of $C_{60}$ involved in the formation of bonds with the Al-surface, we denote length of those four carbon-aluminium bonds by  $ C1$, $C1'$, $C2$, $C2'$. Rolling motion of $C_{60}$ on the Al-surface changes bond lengths of two of these four carbon-aluminium bonds and we denote length of those two carbon-aluminium bonds by  $ C1$ and $C1'$. Rolling motion of $C_{60}$ on the Al-surface does not change bond lengths of other two carbon-aluminium bonds and we denote length of those two carbon-aluminium bonds by  $ C2$ and $C2'$. First we consider the rolling motion of $C_{60}$ molecule on the aluminium surface, where both in the initial and final states, all four carbon atoms forming bonds with the aluminium surface belongs to the pentagonal ring. The energy of the whole system increases for the initial state with the increase in bond length of first two C-Al bonds {\it i.e.}, $C1, C1'$ as given in Table 2. We fix the bond lengths of middle two C-Al bonds {\it i.e.}, $C2, C2'$. Now we look at the decrease in energy of the whole system as a result of formation of two new C-Al bonds $C3, C3'$ {\it i.e.}, for the final state. As we decrease the value of C3, C3' energy of the final state is decreases. We find when (values of C1,C1';,C3,C3') the energy of the initial state is equal to the energy of the final state {\it i.e.}, the transition state, these value of C1,C1';,C3,C3' is used to calculate the energy of transition state.  We find the activation energy for this event by subtracting the energy of the initial state from the energy of the Transition State.

\begin{figure}[ht]
\centering
\begin{subfigure}{.3\textwidth}
\includegraphics[scale=.25]{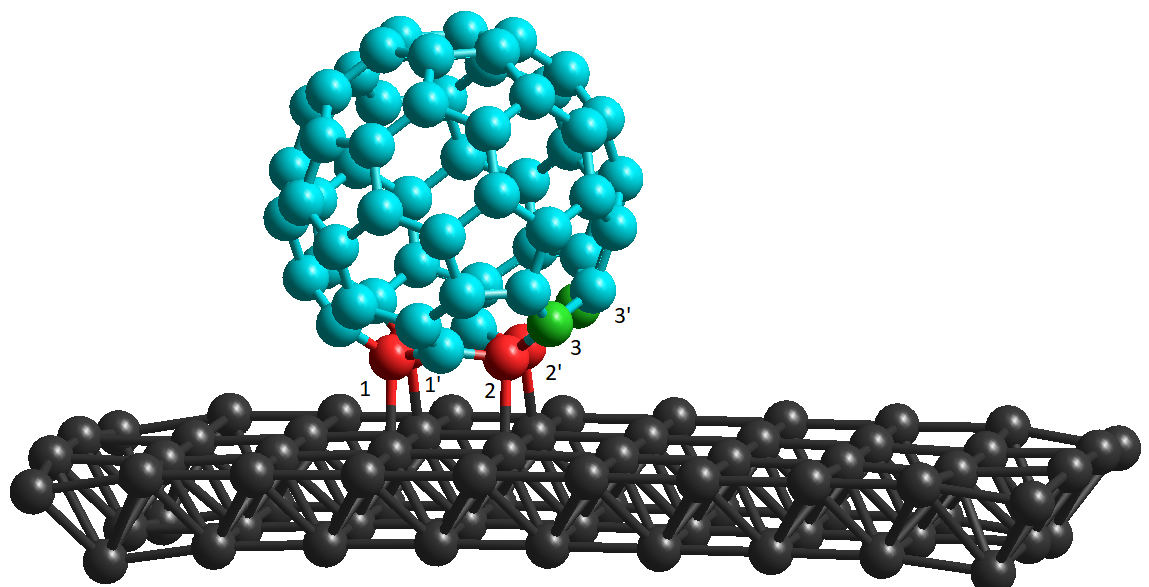}
\caption{Initial state, where all four C-Al bond lengths are equal.}
\end{subfigure}
\begin{subfigure}{.3\textwidth}
\includegraphics[scale=.25]{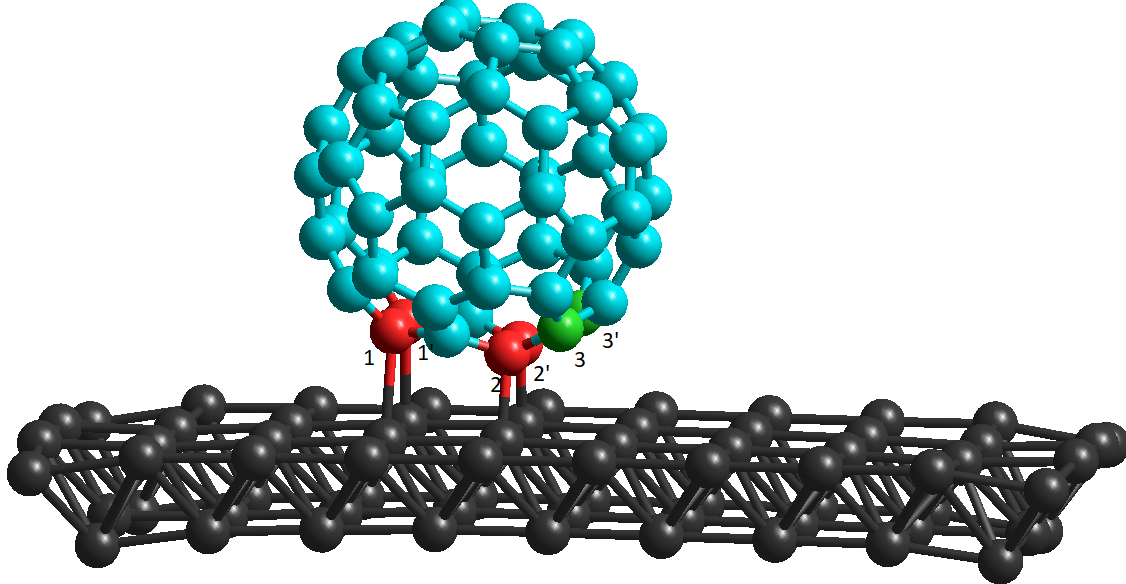}
\caption{Intermediate State, where two C-Al bonds length increases significantly}
\end{subfigure} 
\begin{subfigure}{.3\textwidth}
\includegraphics[scale=.25]{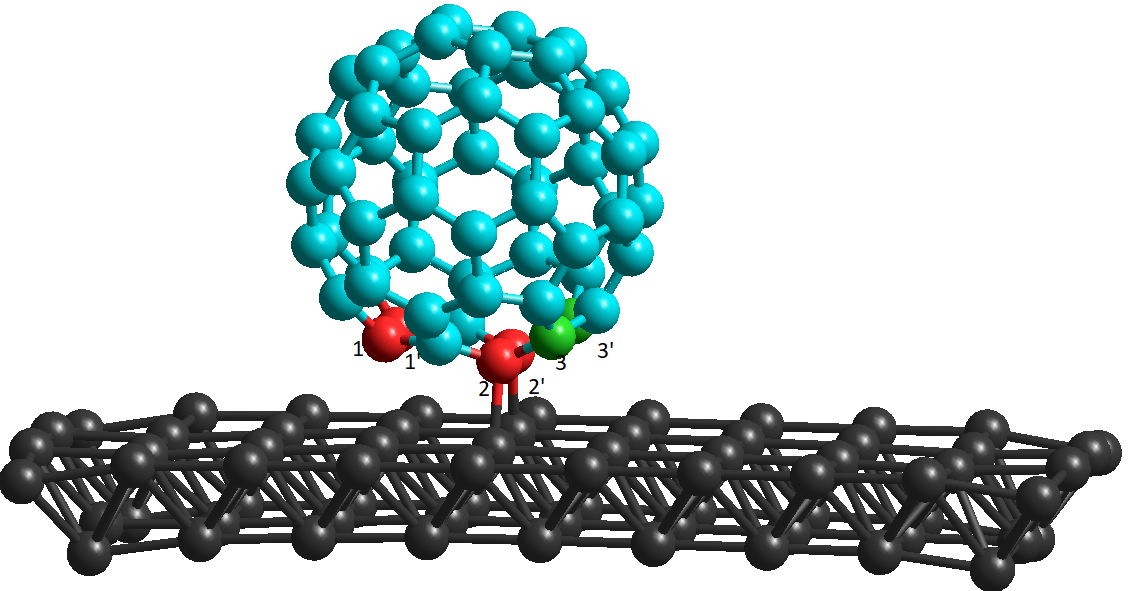} 
\caption{Transition State.}
\end{subfigure}
\begin{subfigure}{.3\textwidth}
\includegraphics[scale=.25]{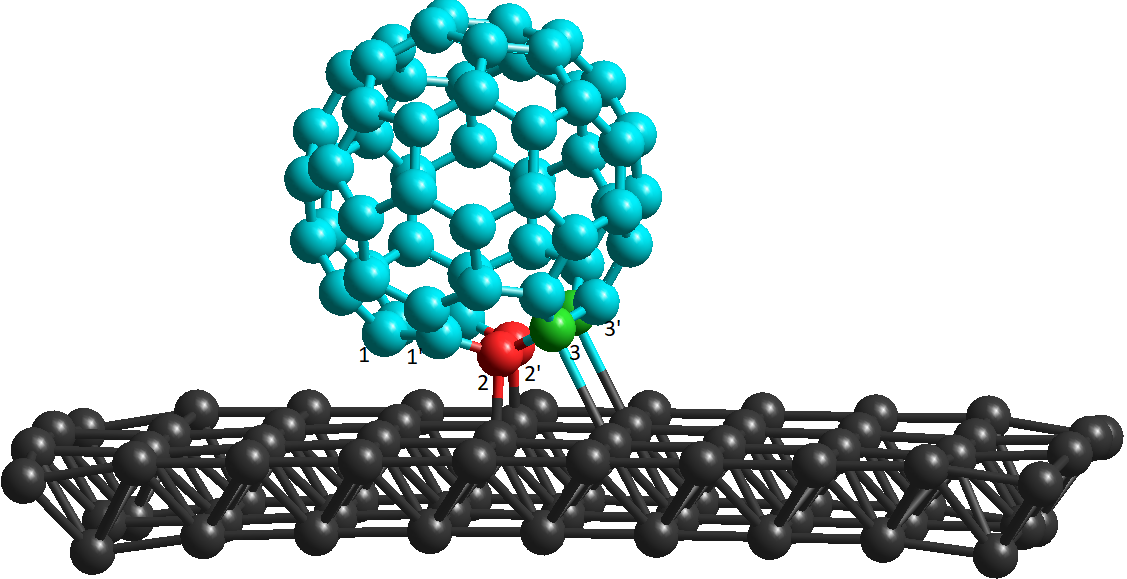}
\caption{Intermediate State, where one can see the possibility of forming two new C-Al bonds.}
\end{subfigure}
\begin{subfigure}{.3\textwidth}
\includegraphics[scale=.25]{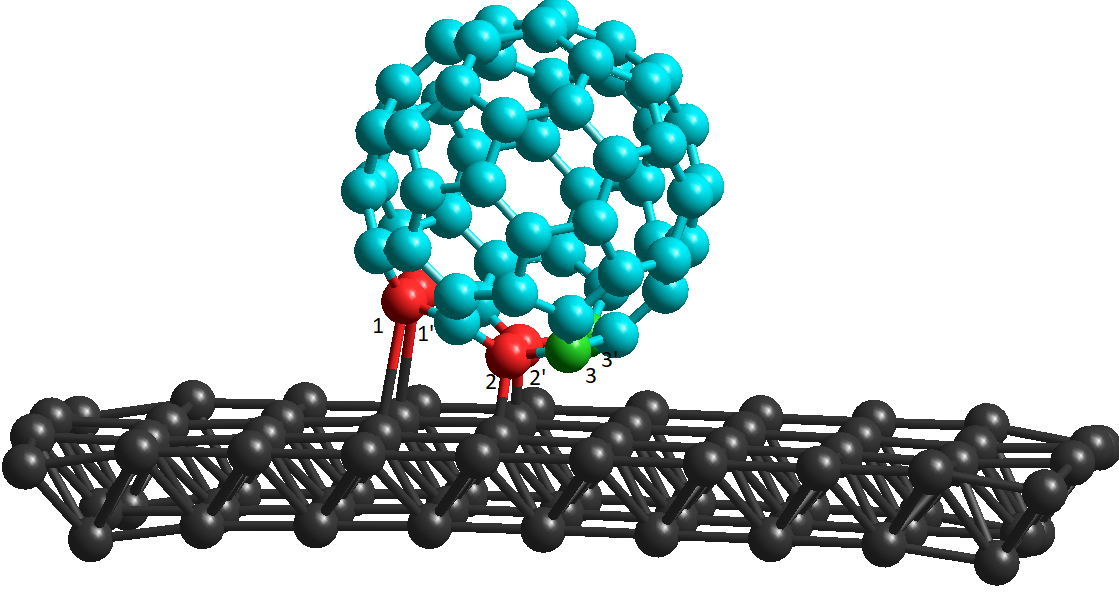}
\caption{Intermediate State, where one can see the breaking two old C-Al bonds.}
\end{subfigure}
\begin{subfigure}{.3\textwidth}
\includegraphics[scale=.25]{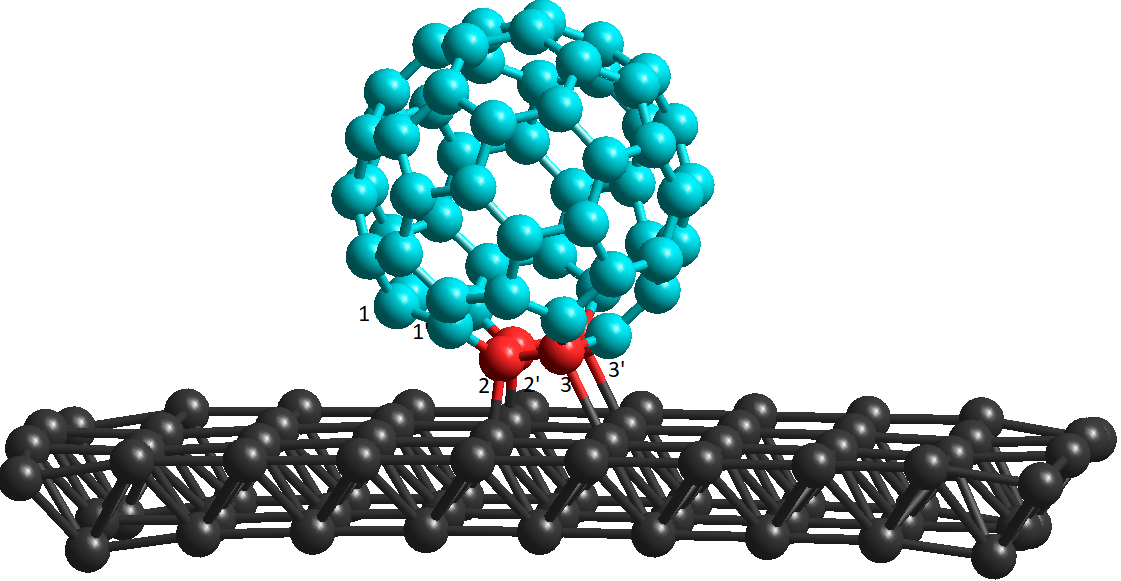}
\caption{Final State, where two new C-Al bonds are formed and two old C-Al bonds are broken.}
\end{subfigure}
\caption{Rolling motion of $C_{60}$ on the surface of Al lattice.}
\end{figure}

%\begin{center}

%\noindent\textbf{Rolling motion from pentagon ring to pentagon ring: }

%\end{center}
\begin{table}
\caption{Rolling motion of $C_{60}$ molecule on the aluminium surface, where both in the initial and final states, all four carbon atoms forming bonds with the aluminium surface belongs to the pentagonal ring. Energy of different states are reported for different values of $C1$, $C1'$, $C3$ and $C3'$.}
\begin{center}
\begin{tabular}{|c|c|c|}
    \hline
    I. S. (kcal mol$^{-1}$) & T.S. (kcal mol$^{-1}$) &  F.S. (kcal mol$^{-1}$)  \\
    \hline
    1112.78 & 1196.12 & 2514.72 \\
    \hline
    1133.71 & 1188.99 & 2196.44 \\
    \hline
    1200.85 & 1179.14 & 1755.09 \\
    \hline
    1284.57 & 1191.62 & 1547.54 \\
    \hline
    1363.84 & 1200.33 & 1434.56 \\
    \hline
    1431.59 & 1211.56 & 1349.85 \\
    \hline
    1507.58 & 1210.31 & 1312.83 \\
    \hline
\end{tabular} 
\end{center}
\end{table}

%\includegraphics[width=8cm]{1.png}
%\hspace{.20cm}
%\includegraphics[width=8cm]{2.png}
%\begin{center}
%    \caption\small{Fig. 6: between bond length1, 2 and Single point Energy For Initial state and For Final state. }
%\end{center}

%\vspace{.30cm}

%\begin{center}
%\begin{tabular}{|c|c|c|c|}
%    \hline
%    Bond & I.S. angle & T.S. angle & F.S. angle  \\
%    \hline
%    bond 1 & \ang{73.83} & \ang{103.09} & \ang{89.65} \\
%    \hline
%   bond 2 & \ang{73.90} & \ang{103.173} & \ang{89.72} \\
%    \hline
%\end{tabular}

%\vspace{.30cm}

%\caption\small{Table 3. Angle between the different Al-C bond and Al-surface of that state when the Bonds are shifts to the next possibility.}

%\end{center}

\noindent In Table 2, we have data for rolling motion of $C_{60}$ molecule on the aluminium surface, where both in the initial and final states, all four carbon atoms forming bonds with the aluminium surface belongs to the pentagonal ring. In the first column of Table 2, we have data for the energy of the initial state for different values of $C1$, $C1'$, $C3$ and $C3'$, in the second column of Table 2, we have data for the energy of the Transition state for different values of $C1$, $C1'$, $C3$ and $C3'$ and in the third column of Table 2, we have data for the energy of the Final state for different values of $C1$, $C1'$, $C3$ and $C3'$. We use those values of $C1$, $C1'$, $C3$ and $C3'$ which are appropriate for rolling motion of $C_{60}$ on the surface of Al lattice. As the the values of $C1$ and $C1'$ increases (bond breaking process), the energy of initial state increases and as the values of $C3$ and $C3'$ decreases (bond making process), the energy of the final state decreases. The value of $C1$, $C1'$, $C3$ and $C3'$ at which energy of the initial state increase to a value which is more than that of the final state - that geometry is called transition state geometry. Using this value of $C1$, $C1'$, $C3$ and $C3'$ one finds the energy of the transition state to be 1211.56 kcal mol$^{-1}$ (from Table 2). Therefore activation energy is estimated by subtracting the energy of the initial state from the energy of the transition state {\it i.e.},  $1211.56 - 1090.18 = 121.38$ kcal mol$^{-1}$. Using a similar method we calculate the energy of the transitions state for rolling motion of $C_{60}$ molecule on the aluminium surface where (i) in the initial state, all four carbon atoms forming bonds with the aluminium surface belongs to the hexagonal ring - in the final state all four carbon atoms forming bonds with the aluminium surface belongs to the pentagonal ring  and  (ii) both in the initial and final states, all four carbon atoms forming bonds with the aluminium surface belongs to the hexagonal ring. The energy of the transition states for case (i) is 1115.57 kcal mol$^{-1}$ and for case (ii) is 1111.24 kcal mol$^{-1}$.

\begin{table}
\caption{Rolling motion of $C_{60}$ molecule on the aluminium surface, where in the initial state all four carbon atoms forming bonds with the aluminium surface belongs to the hexagonal ring and in the final state all four carbon atoms forming bonds with the aluminium surface belongs to the pentagonal ring. Energy of different states are reported for different values of $C1$, $C1'$, $C3$ and $C3'$.}
\begin{center}
\begin{tabular}{|c|c|c|}
    \hline
    I.S. (kcal mol$^{-1}$) & T.S. (kcal mol$^{-1}$) & F.S. (kcal mol$^{-1}$) \\
    \hline
    1125.72 & 1183.07 & 1961.14 \\
    \hline
    1157.63 & 1148.83 & 1815.12 \\
    \hline
    1281.38 & 1126.19 & 1636.90 \\
    \hline
    1379.78 & 1126.19 & 1579.39 \\
    \hline
    1487.17 & 1116.17 & 1497.74 \\
    \hline
    1546.70 & 1115.57 & 1459.75 \\
    \hline
\end{tabular}
\end{center}
\end{table}

%\includegraphics[width=8.5cm]{3.png}
%\hspace{.20cm}
%\includegraphics[width=8.5cm]{4.png}
%\begin{center}
%\caption\small{Fig. 7: between Bond length1, 2 and single point Energy for Initial state and For final state.}

%\end{center}

%\vspace{.20cm}

%\begin{center}
%\begin{tabular}{|c|c|c|c|}
%    \hline
%    Bond & I.S. angle & T.S. angle & F.S. angle  \\
%    \hline
%    bond 1 & \ang{86.616} & \ang{90.795} & \ang{115.617} \\
%    \hline
%   bond 2 & \ang{86.60} & \ang{90.780} & \ang{115.621} \\
%    \hline
%\end{tabular}

%\vspace{.40cm}

%\caption\small{Table 5. Angle between the $C_{60}$ and Al-surface of that state when the bonds are shifts to the next possibility.}
%\end{center}

%\noindent\textbf{Rolling motion from hexagon ring to hexagon ring: }
\begin{table}
\caption{Rolling motion of $C_{60}$ molecule on the aluminium surface, where both  in the initial and final states, all four carbon atoms forming bonds with the aluminium surface belongs to the hexagonal ring. Energy of different states are reported for different values of $C1$, $C1'$, $C3$ and $C3'$.}
 \begin{center}
 \begin{tabular}{|c|c|c|}
    \hline
    I.S.(kcal mol$^{-1}$) & T.S. (kcal mol$^{-1}$) & F.S. (kcal mol$^{-1}$) \\
    \hline
    1123.17 & 1187.96 & 3185.12 \\
    \hline
    1162.20 & 1144.45 & 2792.91 \\
    \hline
    1196.29 & 1134.22 & 2652.47 \\
    \hline
    1270.46 & 1123.97 & 2473.89 \\
    \hline
    1393.44 & 1115.30 & 2299.81 \\
    \hline
    1467.28 & 1111.69 & 2188.70 \\
    \hline
    1561.15 & 1109.44 & 2041.19 \\
    \hline
    1690.67 & 1109.83 & 1868.37 \\
    \hline
    1734.57 & 1110.41 & 1822.25 \\
    \hline
    1823.023 & 1111.24 & 1760.85 \\
    \hline
\end{tabular} 
\end{center}
\end{table}

%\begin{center}
%\includegraphics[width=8cm]{55.png}
%\hspace{.20cm}
%\includegraphics[width=8cm]{6.png}
%\caption\small{Fig. 8: Between bond length1, 2 and single point energy for Initial state and for final state.}
%\end{center}

%\begin{tabular}{|c|c|c|c|}
%    \hline
%    Bond & I.S. angle & T.S. angle & F.S. angle  \\
%    \hline
%    bond 1 & \ang{85.77} & \ang{91.16} & \ang{93.30} \\
%    \hline
%   bond 2 & \ang{85.81} & \ang{91.26} & \ang{93.42} \\
%    \hline
%\end{tabular}
%\vspace{.60cm}
%\caption\small{Table 7. Angle between the $C_{60}$ and al-surface of that state when the bonds are shifts to the next possibility.}

\noindent We find the activation energy for the rolling motion of $C_{60}$ molecule on the aluminium surface, where (a) both  in the initial and final states, all four carbon atoms forming bonds with the aluminium surface belongs to the hexagonal ring (with imaginary frequency 3.16 $cm^{-1}$ ), (b) in the initial state, all four carbon atoms forming bonds with the aluminium surface belongs to the hexagonal ring and in the final state  all four carbon atoms forming bonds with the aluminium surface belongs to the pentagonal ring (with imaginary frequency 25.62 $cm^{-1}$) and (c)  both  in the initial and final states, all four carbon atoms forming bonds with the aluminium surface belongs to the pentagonal ring (with imaginary frequency 57.73 $cm^{-1}$ ) and we report activation energy values for all three cases in Table 8. Normal model analysis is performed and the rate constants are estimated using classical version of Transition state theory and we report rate constant values at room temperature for all three cases in Table 8.

\section{Conclusions}
\noindent It is easy for the $C_{60}$ molecule to roll on the surface of Al lattice when both in the initial and final states, all four carbon atoms forming bonds with the aluminium surface belongs to the hexagonal ring and the rate constant of this process is very fast with rate constant $3.45 \times 10^{13}$. Therefore rolling motion of $C_{60}$ molecule on the surface of Al lattice continues through all states where  all four carbon atoms forming bonds with the aluminium surface belongs to the hexagonal ring. It is also possible for the $C_{60}$ molecule to roll with a slower speed on the surface of Al lattice when in the initial state, all four carbon atoms forming bonds with the aluminium surface belongs to the hexagonal ring and in the final state  all four carbon atoms forming bonds with the aluminium surface belongs to the pentagonal ring.

\begin{table}
\caption{Activation energy and rate constant for different types of rolling motion.}
\begin{center}
\begin{tabular}{|c|c|c|}
   \hline
   Rolling with bonds  & $E_{act}$ (kcal mol$^{-1}$) & Rate const. $sec^{-1}$ \\
   \hline
   Hexagon and Hexagon  & 1.69 & $3.45 \times 10^{13}$  \\
   \hline
   Hexagon and Pentagon & 5.97 & $28.92 \times 10^{2}$ \\
   \hline
   Pentagon and Pentagon & 121.38 & $17.01 \times 10^{-79}$\\
   \hline
\end{tabular} 
\end{center}
\end{table}

\section{Acknowledgements}

\noindent One of the authors (S.K.) thanks Dr. Bidisa Das for several interesting discussions and another author (A.C.) thanks IIT Mandi for providing PDA. This work is a part of S.K.'s M.Sc. (chemistry) thesis.

\end{document}